\documentclass[journal,twoside,web]{ieeecolor}
\usepackage{generic}
\usepackage{cite}
\usepackage{amsmath,amssymb,amsfonts}
\usepackage{algorithmic}
\usepackage{graphicx}

\usepackage[hidelinks]{hyperref} 
\usepackage{textcomp}
\def\BibTeX{{\rm B\kern-.05em{\sc i\kern-.025em b}\kern-.08em
    T\kern-.1667em\lower.7ex\hbox{E}\kern-.125emX}}
\begin{document}
\title{On the Performance of Dual-Gate Reconfigurable Nanowire Transistors}
\author{Bin Sun, Benjamin Richstein, Patrick Liebisch, Thorben Frahm, Stefan Scholz, Jens Trommer, \IEEEmembership{Member, IEEE}, Thomas Mikolajick, \IEEEmembership{Senior Member, IEEE}, Joachim Knoch, \IEEEmembership{Senior Member, IEEE}
\thanks{This work was partially supported by Deutsche Forschungsgemeinschaft (DFG, German Research Foundation) under Germany’s Excellence Strategy -
Cluster of Excellence Matter and Light for Quantum Computing (ML4Q) EXC 2004/1 - 390534769, under KN545/22-1 and KN545/29-1. The review of this article was arranged by Editor xx. \textit{(Corresponding author: Joachim Knoch.)}}
\thanks{B. Sun, B. Richstein, P. Liebisch, T. Frahm, S. Scholz, and J. Knoch are with the Institute of Semiconductor Electronics, RWTH Aachen University, 52056 Aachen, Germany (e-mail: sun@iht.rwth-aachen.de; knoch@iht.rwth-aachen.de).}
\thanks{J. Trommer is with the NaMLab gGmbH, 01069 Dresden, Germany}
\thanks{T. Mikolajick is with NaMLab gGmbH, 01187 Dresden, Germany, also
with the Chair of Nanoelectronic Materials, TU Dresden, 01187 Dresden,
Germany, and also with the Center for Advancing Electronics Dresden
(CfAED), TU Dresden, 01062 Dresden, Germany}
}

\maketitle

\begin{abstract}
We investigate the operation of dual-gate reconfigurable field-effect transistor (RFET) in the program-gate at drain (PGAD) and program-gate at source (PGAS) configurations. To this end, dual-gate silicon nanowire (SiNW) FETs are fabricated based on anisotropic wet chemical silicon etching and nickel silicidation yielding silicide-SiNW Schottky junctions at source and drain. Whereas in PGAD-configuration ambipolar operation is suppressed, switching is deteriorated due to the injection through a Schottky-barrier. Operating the RFET in PGAS-configuration yields a switching behavior close to a conventional MOSFET. This, howewer, needs to be traded off against strongly  non-linear output characteristics for small bias.  
\end{abstract}


\section{Introduction}
\label{sec:introduction}
\IEEEPARstart{I}{n} recent years, device scaling close to the physical limit 
has spurred the move from geometrical to equivalent scaling \cite{IRDS} of conventional MOSFETs utilizing new device architectures, materials and integration schemes to continue delivering integrated circuits with higher density 
and improved performance (power, speed). In addition, fabrication techniques such as ion implantation, yielding \textit{n-} as well as \textit{p-}type conduction of an intrinsic semiconductor, gradually become less suitable as device structures are being aggressively miniaturized into the deep nanoscale. The finite solid solubility of dopants and their increased ionization energy in nanoscale dimensions \cite{Bjork2009} could lead to increased parasitic resistances and capacitances in the source/drain regions. Furthermore, the statistical nature of ion implantation and dopant diffusion during the activation anneal make it hard to implement well-defined potential profiles with abrupt interface transitions within a small volume of semiconductor material, leading to a variability in threshold voltage $V_\text{th}$ from device to device due to random dopant distribution effects. Moreover, the potential landscape created by physical doping stays fixed after the implantation/activation process which only allows either \textit{n}- or \textit{p}-type unipolar operation mode. 

The aforementioned issues are best addressed if doping in source/drain and channel can be avoided altogether. Recently, reconfigurable FETs (RFETs)  have attracted an increasing attention because they utilize electrostatic doping to create virtual \textit{n}-/\textit{p}- regions avoiding dopant related issues. Moreover, they can be tuned to operate as \textit{n}-/\textit{p}-transistors with unipolar device operation similar to conventional MOSFETs \cite{6479004,9099545,Heinzig2012,7903719,mueller2016}. Previous work using tri-gates \cite{6479004,mueller2016}, i.e. two for creating virtual source/drain regions and the third for on/off switching, has demonstrated reconfigurable devices with high $I_\text{on}$/$I_\text{off}$ ratio and steep switching behavior. The adoption of three gates will certainly need more area and might become a key limiting factor for down-scaling. Hence, devices using only a dual-gate architecture have been investigated \cite{9099545,Heinzig2012,7903719}. In these devices, the gate at the drain contact is used to \lq program\rq\, the device either to \textit{n-} or \textit{p-}type operation while the gate at the source is used as the switching or control gate. However, while unipolar device operation is obtained with this program-gate at drain configuration (PGAD), the inverse subthreshold slope is significantly larger than 60\,mV/dec. This is obvious since the RFET in PGAD configuration is simply a Schottky-barrier MOSFET with suppressed ambipolar operation \cite{TED2006}. In this work, we investigate the device configuration with the program gate at source (PGAS) and compare it with the PGAD. As it turns out, the PGAS configuration yields a switching behavior similar to a conventional MOSFET with the drawback of a larger off-state leakage. Furthermore, PGAD and PGAS show a  distinctly different non-linearity in the output characteristics with a stronger effect in the PGAS configuration. 
\begin{figure}[!t]
\centerline{\includegraphics[width=\columnwidth]{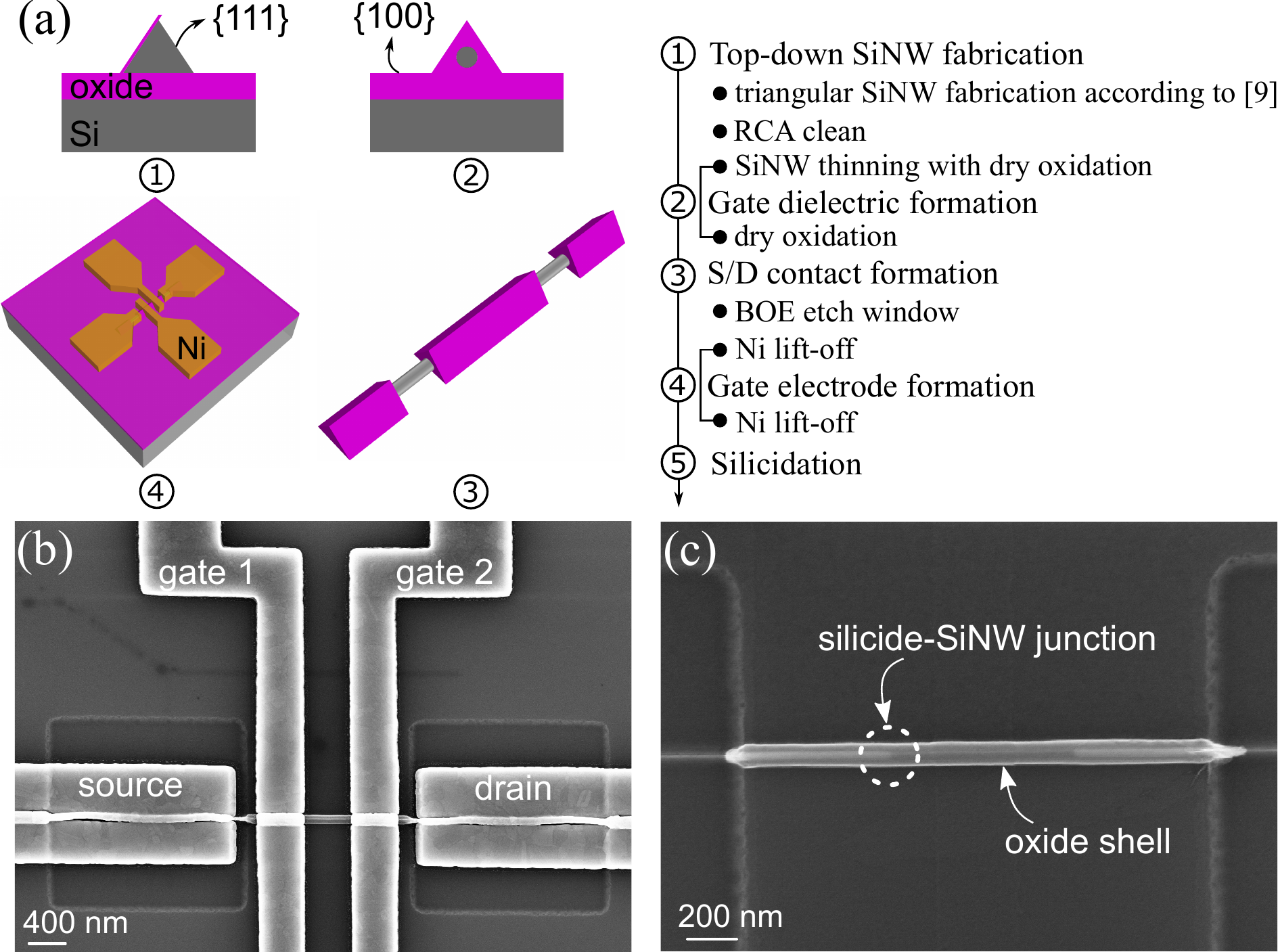}}
\caption{Schematics (dimension is not to scale) and key process steps in device fabrication (a). Scanning electron micrographs of the final device after Ni lift-off and silicidation processes (b) and after selective removal of unreacted Ni (including gate electrodes) (c).}
\label{fig:sun1}
\end{figure}
\section{Device Fabrication}
\label{sec:device_fabrication}
Reconfigurable silicon nanowire FETs are fabricated using a top-down approach with a two-step wet chemical etching as described in \cite{doi:10.1063/1.4737463}. 
Tetramethylammonium hydroxide (TMAH) instead of potassium hydroxide (KOH) is chosen as the wet chemical etchant of silicon to ensure a fully CMOS-compatible fabrication process. A \{111\} facet inclined at 54.74\,$^{\circ}$ with respect to the \{100\} facet is formed due to the distinct etch-rate anisotropy of different crystallographic planes \cite{MERLOS1993737} in an alkaline solution (i.e. TMAH) when a silicon nitride hardmask is defined in a line-pattern along the [110] direction of a (100)-oriented silicon substrate. Exploiting the ability of SiN to act as effective diffusion barrier enables a local oxidation of the etched \{111\} facet. After selective removal of the SiN a triangle bounded by two \{111\} facets is formed in a second TMAH etching since the \{111\} facet fabricated with the first TMAH etch step is protected by the grown, local oxide (cf. Fig.~\ref{fig:sun1}(a), step 1). Atomic force microscopy is used to probe the surface roughness of the wet-etched \{111\} facet formed on a step structure (step height $\sim$\,800\,nm) in a bulk silicon substrate, the measured root-mean-square (RMS) value = 0.3992\,nm from a rectangular scanning area (800\,nm$\times$250\,nm) is well within the range of a commercial prime-grade silicon wafer (RMS = 0.2–0.8\,nm) \cite{kemme}. If such a process is carried out on a (100)-oriented silicon-on-insulator substrate (\textit{p}-type boron doped, 10$^{\text{15}}$\,cm$^{\text{-3}}$, top-Si $\sim$\,80\,nm, buried oxide $\sim$\,145\,nm), the buried-oxide acts as an etch stop layer. Hence the size of the triangle mainly depends on the thickness of the top silicon layer. Subsequently, a dry oxidation step is carried out to reduce the size of the SiNW and form a gate oxide at the same time. The dry oxidation is carried out  for 3\,min at a temperature of 1100\,$^{\circ}$C, i.e. higher than the viscous flow point. Although such a high temperature prevents a self-limiting oxidation \cite{doi:10.1116/1.2198847} it allows transforming the cross-section of the SiNW from triangular to a circular shape while minimizing the degradation of the Si/SiO$_\text{2}$ interface due to roughness induced by the oxidation process \cite{Fang_1997}. Next, etch windows are patterned in the source/drain contact areas to remove the oxide shell with buffered oxide etch (BOE). Subsequently, nickel contacts and gates are fabricated with electron beam lithography using a bilayer PMMA resist, electron beam evaporation and lift-off. Finally, the sample is annealed in forming gas (mixture of 90\,\% N$_\text{2}$ and 10\,\% H$_\text{2}$) at 450\,$^{\circ}$C to silicide the SiNW and drive the silicide-SiNW junction underneath the two gate electrodes to avoid a gate underlap and achieve a good electrostatic control of the silicide-SiNW junctions; Fig.~\ref{fig:sun1}(c) shows this Ni intrusion into a SiNW surrounded by an oxide shell after the selective removal of the unreacted Ni in source/drain regions (which also removes the two gates). Fig.~\ref{fig:sun1}(b) shows a scanning electron micrograph of a readily fabricated reconfigurable dual-gate FET.

\section{Device Characterization}
\label{sec:device_characterization}
$I$-$V$ characteristics of the reconfigurable field-effect transistors are measured in PGAD- and PGAS-configurations and compared with each other. 
In both (PGAS and PGAD) configurations, a constant voltage applied to the PG generates a virtual \textit{p}- or \textit{n}-type doping section whereas the other gate - called control gate (CG) - is used to switch the transistor.

\begin{figure}[!t]
\centerline{\includegraphics[width=\columnwidth]{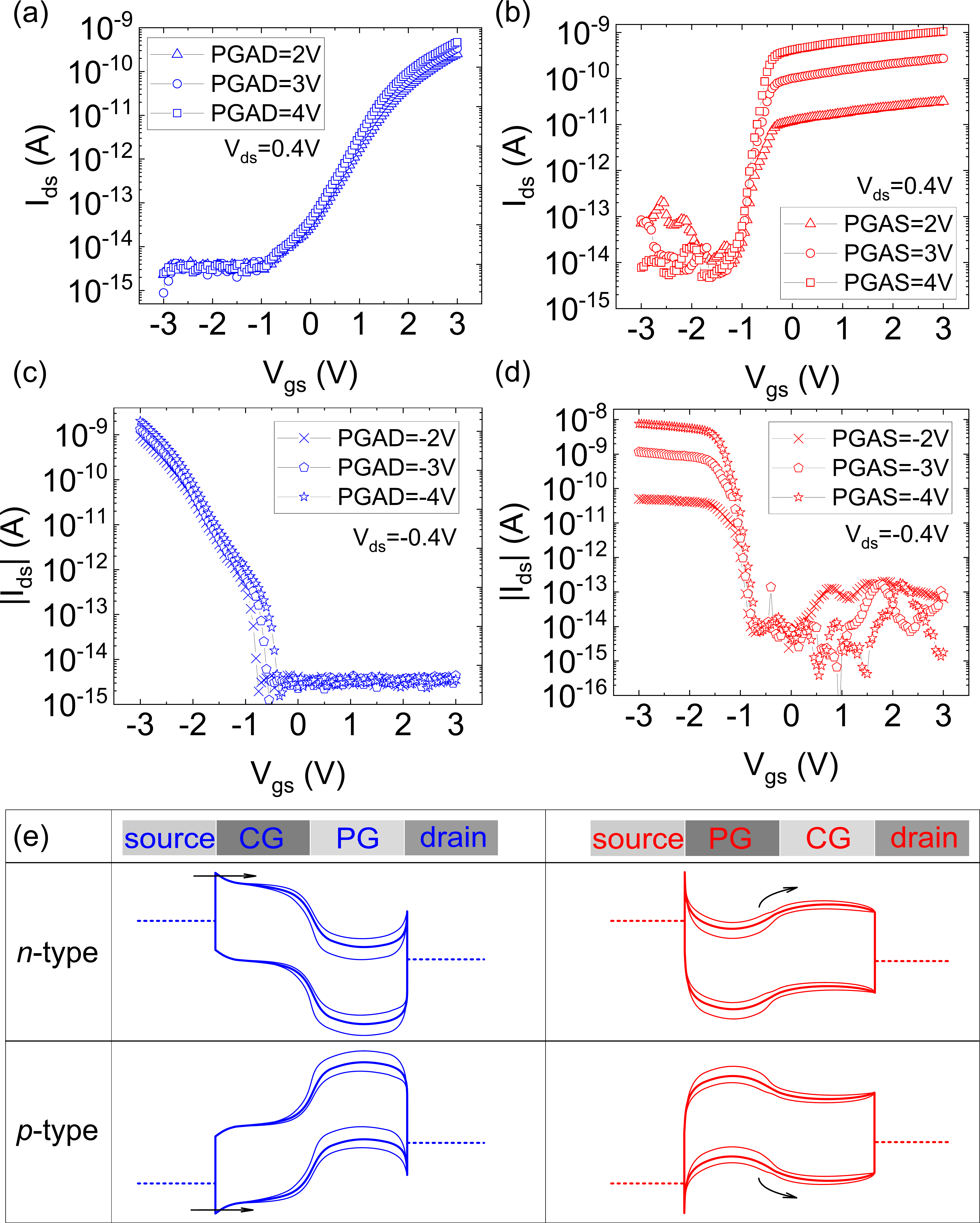}}
\caption{Transfer characteristics of a RFET for different program gate voltages in PGAD configuration for \textit{n}-type operation mode (a), in PGAS configuration for \textit{n}-type operation mode (b), in PGAD configuration for \textit{p}-type operation mode (c), and in PGAS configuration for \textit{p}-type operation mode (d), and their corresponding band diagrams (e). Gate leakage currents (\unboldmath{$I_\text{CG}$, $I_\text{PG}$)} are below $\sim$\,5$\times$10$^{\text{-15}}$\,A and are not plotted for clarity.}
\label{fig:sun2}
\end{figure}
\begin{figure}[!t]
\centerline{\includegraphics[width=\columnwidth]{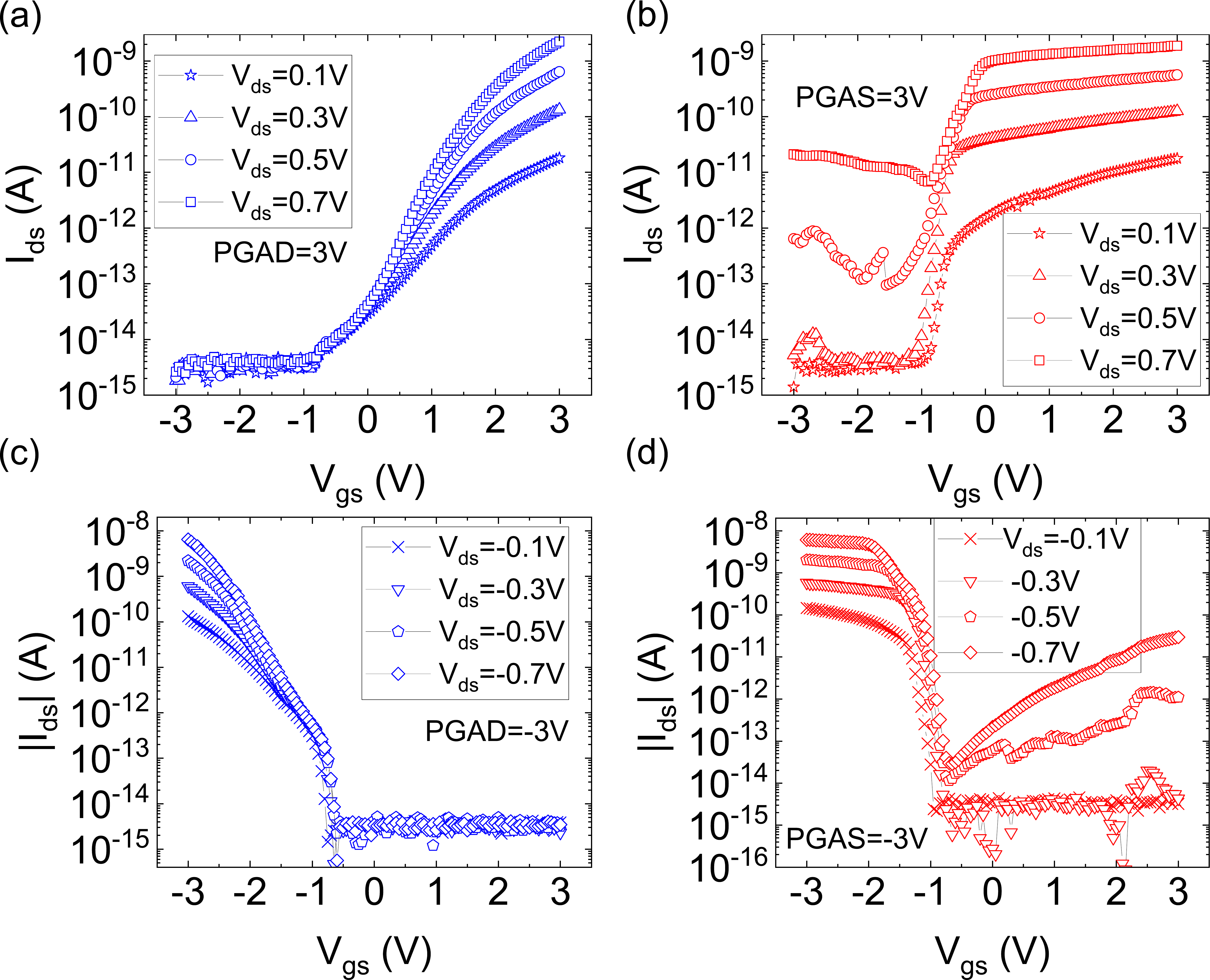}}
\caption{Transfer characteristics of the same device for different drain source voltages in PGAD configuration for \textit{n}-type operation mode (a), in PGAS configuration for \textit{n}-type operation mode (b), in PGAD configuration for \textit{p}-type operation mode (c), and in PGAS configuration for \textit{p}-type operation mode (d).}
\label{fig:sun4}
\end{figure}

\begin{figure}[!t]
\centerline{\includegraphics[width=\columnwidth]{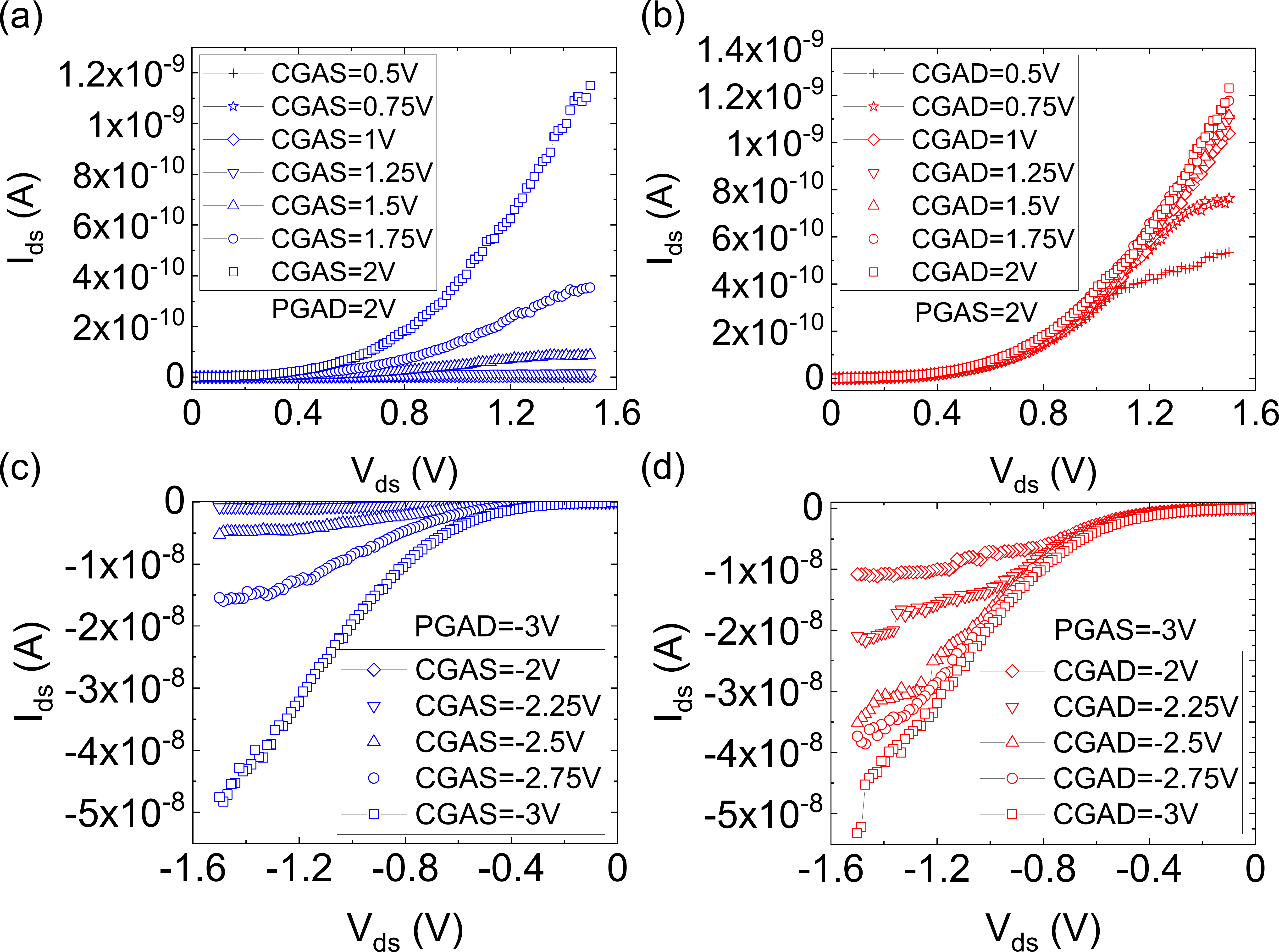}}
\caption{Output characteristics of the same device for different control gate voltages in PGAD configuration for \textit{n}-type operation mode (a), in PGAS configuration for \textit{n}-type operation mode (b), in PGAD configuration for \textit{p}-type operation mode (c), and in PGAS configuration for \textit{p}-type operation mode (d).}
\label{fig:sun3}
\end{figure}


Figure~\ref{fig:sun2} shows transfer curves of a RFET in PGAS (red curves in the right panels) and PGAD configurations (blue curves in the left panels) for both \textit{n-} and \textit{p-}type operation modes with increasing $V_\text{PG}$ values and their corresponding band diagrams. In the case of PGAS, a virtual \textit{n}-doped source contact is created at the source side if a positive voltage is applied to the program gate, 
and the behavior of the device is similar to a conventional \textit{n}-type MOSFET in terms of a steep inverse subthreshold slope. The on-state current increases with a larger $V_\text{PG}$ due to the increased carrier injection at the source Schottky junction. In PGAD configuration, a virtual \textit{n}-type drain contact is created if a positive voltage is applied to the program gate. 
In this case, the carrier injection through the source-side Schottky-barrier is modulated by different $V_\text{CG}$ which makes the transistor behave similar to a conventional Schottky-barrier MOSFET. Therefore, the on-state current mainly remains constant even if a larger $V_\text{PG}$ is applied at the drain side. The transition of carrier injection from thermionic emission to tunneling through the Schottky-barrier at the source side is reflected in a kink in the transfer characteristic in Fig.~\ref{fig:sun2}(c). Such a kink is not visible for the \textit{n}- branch in Fig.~\ref{fig:sun2}(a), which means that the Schottky-barrier for hole injection is smaller than that for electrons.\\
\indent Figure~\ref{fig:sun4} shows transfer curves of the same device in PGAS and PGAD configurations for both \textit{n-} and \textit{p-}type operation modes with increasing $V_\text{ds}$. A current increase enabled by the modulation of the exit resistance at the drain side is observed for all cases. However, while in PGAD configuration the leakage remains below the measurable threshold due to the particular potential profile throughout the device, larger $V_\text{ds}$ values lead to an exponential increase in leakage current in the PGAS configuration. The reason for this is that in PGAS configuration, the ambipolar operation due to an increased carrier injection at the drain Schottky junction with increasing $V_\text{ds}$ is not suppressed.

Figure~\ref{fig:sun3} shows output curves of the same device in PGAS and PGAD configurations for both \textit{n-} and \textit{p-}type operation modes with increasing $V_\text{CG}$ values. In the case of PGAD (i.e. CGAS, blue curves in the left panels), a non-linear current increase in the triode regime is observed, typical of a Schottky-barrier MOSFET. In the PGAS configuration, a non-linear current increase for small bias is also observed. However, in contrast to PGAD, an exponential increase is obtained that is almost the same for all CG voltages displayed in Fig.~\ref{fig:sun3}
 (b) and (d); a saturation of the current at different levels occurs only for $V_\text{ds}$ larger than $\sim$\,1\,V. Both non-linearities, i.e. in PGAD and PGAS, deteriorate the functionality of logic inverters built from those devices. However, the PGAS configuration is particularly detrimental leading to a complete loss of noise margin. 
 
 The different behavior of the output characteristics of the two device configurations may at first be surprising since it is generally argued that the non-linearity of the $I_\text{d}-V_\text{ds}$ curves stems from the forward biased drain Schottky junction. As will become clear below, this is not the case: while the non-linearity of the PGAS configuration is indeed due to a forward-biased drain Schottky-junction, the non-linearity of the PGAD configuration is due to an increased tunneling through the source-side Schottky-barrier. 

\section{Device Simulations}
\label{sec:device_sim}

\begin{figure}[!t]
\centerline{\includegraphics[width=\columnwidth]{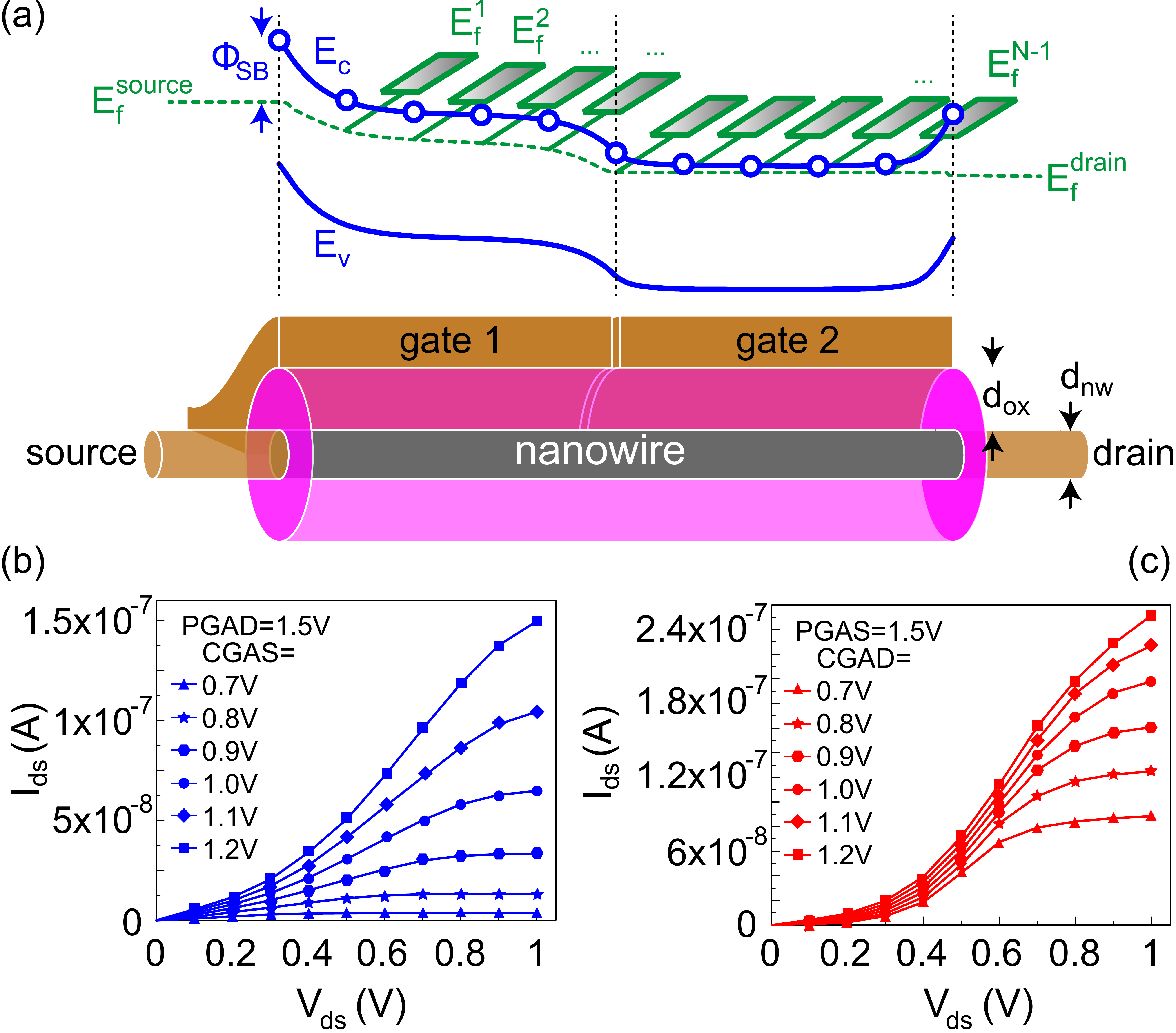}}
\caption{Schematic of the simulated dual-gate nanowire device (a). The top panel shows the conduction and valence bands (blue lines) on a finite difference grid. Buettiker probes are connected to each finite difference site with \unboldmath{$E_\text{f}^\text{i}$} representing the quasi-Fermi level (green dashed line) through the device. Simulated \textit{n-}type output characteristics using NEGF approach for both PGAD (b) and PGAS (c) configurations. For the simulation a device with \unboldmath{$L$}=100\,nm, \unboldmath{$d_\text{nw}$}=1\,nm, \unboldmath{$d_\text{ox}$}=10\,nm, \unboldmath{$E_\text{g}$}=1\,eV, \unboldmath{$m_\text{c,v}^{\star}$} and a scattering mean free path of 15\,nm are assumed.}
\label{fig:sun5}
\end{figure}
In order to elaborate further on the behavior of RFETs and in particular to understand the difference in the non-linear behavior of the PGAS and PGAD configurations, self-consistent device simulations using the non-equilibrium Green's function approach on a finite difference grid have been carried out \cite{datta}. An effective mass approximation with symmetric conduction and valence band is assumed; the complex band structure within the band gap is taken into account with Flietner's dispersion relation \cite{flietner1972}. To reduce the computational burden, a nanowire FET with one-dimensional electronic transport is considered. Furthermore, the electrostatics is described by a 1D modified Poisson equation where the device architecture is accounted for with a proper choice of a natural lengths scale $\lambda$ which represents the characteristic length on which potential variations are being screened \cite{yan,auth}. This simulation approach has been used successfully to reproduce and explain the device behavior of conventional as well as Schottky-barrier field-effect transistors (see e.g. \cite{TED2002,apa}). 

An important ingredient here is the inclusion of Buettiker probes (cf. Fig~\ref{fig:sun5} (a)), i.e. virtual contacts attached to the channel of the RFET that mimic inelastic scattering \cite{Venugopal2003}; the Fermi levels of all Buettiker probes are determined to ensure a net zero current from/into each Buettiker probe thus representing the quasi Fermi level within the device. Fermi level pinning with a midgap position of the source/drain Fermi level yielding a Schottky-barrier of $\Phi_\text{SB}=E_\text{g}/2$ is assumed at the interface between source/drain and the channel; all other device parameters are stated in Fig.~\ref{fig:sun5}. 

\begin{figure}[!t]
\centerline{\includegraphics[width=\columnwidth]{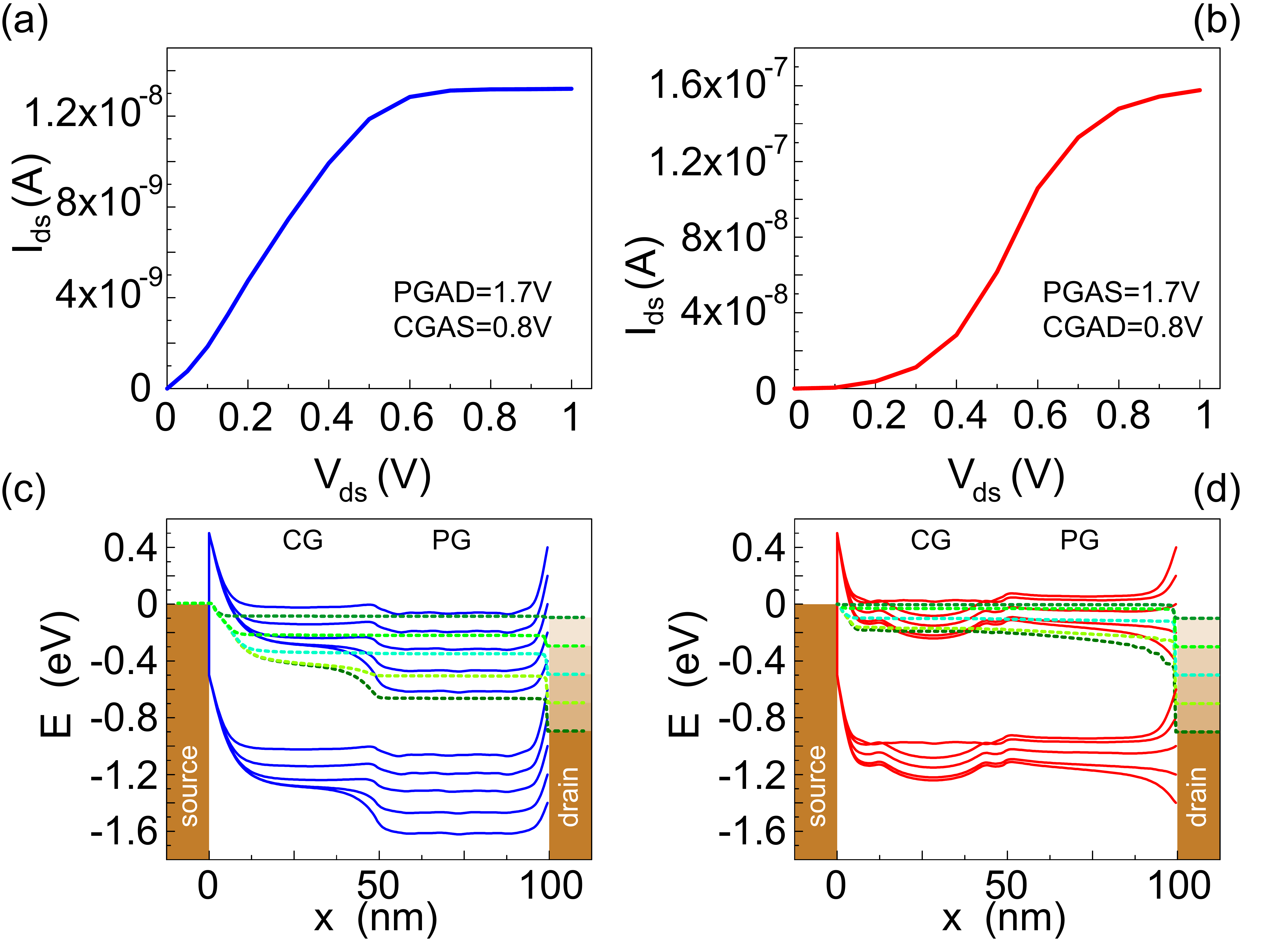}}
\caption{Output characteristic for PGAD=1.7\,V, CGAS=0.8\,V (a) and PGAS=1.7\,V, CGAD=0.8\,V (b). The lower panels show the conduction/valence band profiles for \unboldmath{$V_\text{ds}$}=0.1, 0.3, 0.5, 0.7, 0.9\,V together with the quasi Fermi level in the two device configurations.}
\label{fig:sun6}
\end{figure}
Figure~\ref{fig:sun5} depicts simulated output characteristics in PGAD (b) and PGAS (c) configuration, qualitatively reproducing the experimental results displayed in Fig.~\ref{fig:sun3}. In particular, the strong non-linear current increase for small bias in the PGAS configuration compared to PGAD is observed. 
Figure~\ref{fig:sun6} shows a single $I_\text{d}$-$V_\text{ds}$ curve for PGAD=1.7\,V, CGAS=0.8\,V (a) and PGAS=1.7\,V, CGAD=0.8\,V (b) showing clearly the different behavior of the two device configurations: while PGAD exhibits an almost linear behavior in the region of small bias\footnote{Note, that this is the case due to the large difference in $V_\text{PG}$ and $V_\text{CG}$: For larger CG voltages as displayed in Fig.~\ref{fig:sun5} (b), the output characteristics also become non-linear in the PGAD configuration.}, PGAS shows a strong non-linearity. The reason for this different behavior becomes obvious when comparing Fig.~\ref{fig:sun6} (c) and (d): in PGAD, the quasi Fermi level (green dashed lines) at the drain end, i.e. underneath gate 2, drops substantially with increasing bias. This is due to a loss of carrier injection from drain such that the conduction/valence bands are moved to lower energies because of the large $V_\text{PG}$ at gate 2. As a result, the device does not behave like a forward biased Schottky junction. Instead, the small non-linearity observed in (a) stems from an increased tunneling through the source-side Schottky-diode with increasing $V_\text{ds}$ due to a significant reduction of the carrier density underneath the control gate (cf. Fig.~\ref{fig:sun6} (c)). Note that the same is true in conventional, single-gate Schottky-barrier MOSFETs \cite{nanoelectronicsbook}. In contrast, in the case of PGAS (Fig.~\ref{fig:sun6} (b) and (d)) the quasi Fermi level remains rather fixed by the source contact and increasing the bias yields a forward biased Schottky-junction with an exponential increase of the current. This rather undesirable behavior strongly deteriorates, e.g., the performance of inverters made of RFETs configured in PGAS. 

In conclusion, with only two individually controllable gates, the PGAS configuration yields an almost ideal switching behavior while its output characteristics are deteriorated when compared to PGAD which exhibits improved output but deteriorated transfer characteristics. 

\section{Conclusion}
\label{sec:conclusion}
Reconfigurable FETs with wet chemically etched SiNWs and Ni silicidation are fabricated and characterized for both PGAS and PGAD configurations. The top-down fabrication approach yields localized SiNWs with atomically flat surfaces and minimal plasma damage manufacturable with conventional h-line contact lithography. The one-step silicidation process at 450\,$^{\circ}$C creates a silicide composition that has a Fermi level close to the valance band of silicon. 
In PGAS configuration, the RFET demonstrates a steep inverse subthreshold slope comparable to a state-of-the-art conventional MOSFET. However, $V_\text{ds}$ should be kept low to prevent leakage current increase induced by the other carrier type. In PGAD configuration, the RFET is a Schottky-barrier MOSFET with suppressed ambipolar behavior. The price paid for a lower off-state leakage current and more unipolar device behavior is a deteriorated switching behavior. The quasi Fermi level through the device simulated with Buettiker probes to mimic scattering explains the difference in output characteristics for both PGAS and PGAD configurations.  



\section*{Acknowledgment}
Bin Sun is grateful for financial support from the China Scholarship Council. The authors thank M. Liebmann (2nd Institute of Physics B, RWTH Aachen University) for carrying out AFM measurements.
\bibliographystyle{IEEEtran}
\bibliography{IEEEabrv,bibliography}

\end{document}